\documentclass{prop2015}
\usepackage[english]{babel}
\def\be{\begin{equation}}
\def\ee{\end{equation}}
\def\bea{\begin{eqnarray}}
\def\eea{\end{eqnarray}}
\keywords{Romans Supergravity, Massive supergravities, M-theory, Supermembrane, M2-brane.}
\title{10D Massive Type IIA Supergravities as the uplift of Parabolic M2-brane Torus bundles}
\author[Maria Pilar Garcia del Moral]{Maria Pilar Garcia del Moral\inst{1}\footnote{Maria Pilar Garcia del Moral\quad E-mail:\quad~\textsf{maria.garciadelmoral@uantof.cl}}}
\author[S.\,Alvaro Restuccia]{Alvaro Restuccia\inst{1,2}}
\address[1]{Depto de F\'isica, Universidad de Antofagasta, Aptdo 02800, Chile.}
\address[2]{Depto de F\'isica, Universidad de Antofagasta, Aptdo 02800, Chile; and
Departamento de F\'\i sica, Universidad Sim\'on Bol\'\i var,
Apartado 89000, Caracas 1080-A, Venezuela.}
\shortauthors{M.P. Garcia del Moral et al.}
\begin{abstract}
  We remark that the two 10D massive deformations of the $N=2$ maximal type IIA supergravity  (Romans and HLW supergravity) are associated to the low energy limit of the uplift to 10D of M2-brane torus bundles with parabolic monodromy linearly and  non-linearly realized respectively. Romans supergravity corresponds to M2-brane compactified on a twice-punctured torus bundle.
\end{abstract}
\shortabstract
\begin{document}
\maketitle
\section{Introduction} In ten dimensions there are two different maximal Supergravities (IIA and IIB) which correspond respectively to the low energy limit of the 10D type IIA and IIB String Theories. There are also two massive deformations in the 10D type IIA sector. One of them found in \cite{romans} is Romans Supergravity, a massive supergravity whose covariant 11D Supergravity origin is not known and the second one, that we will denote by HLW  \cite{hlw}, is a gauged supergravity that can be obtained by a Scherk Schwarz reduction(SS)\cite{SS} from 11D Supergravity. The $10D$, $SL(2,R)$  Type IIB supergravity does not contain any massive deformation.  Romans is a genuine massive supergravity theory. It does not admit as a solution a 10D Minkowski background but instead it has Domain walls solutions associated to $D8$ branes \cite{bct}. In \cite{blo} it was shown that its 11D supergravity uplift is related to the existence of a noncovariant cosmological term. There is a well-known no-go theorem \cite{deser} emphasizing the inexistence of a 11D covariant deformation 11D supergravity. It can be evaded at the price of introducing an isometry with an associated killing vector that breaks it \cite{blo}. In \cite{hull-massive} it was pointed out that Romans supergravity is the low energy limit of a massive string theory. HLW supergravity on the other hand is a supergravity formulation without lagrangian. It appears as the gauging of the scaling symmetry of equation of motion in 11D, it has been called \textit{trombone symmetry}. The allowed background solution is De Sitter. Both theories when reduced a la Kaluza Klein (KK) to 9D become type II gauged supergravities:  Romans becomes a parabolic gauged supergravity and HLW a gauged trombone supergravity, \cite{tp, bergshoeff, torrente}. \newline
In \cite{gmpr3,joselen} it  has been proposed a new formulation of the Supermembrane theory (M2-brane) on a  symplectic torus bundles with $SL(2,Z)$ monodromy as the M-theory origin of type II gauged supergravity in 9D. It was shown that the supermembrane with central charges is the M-theory origin of type IIB gauged supergravities in 9D. Since it is also a U-dual invariant action\cite{t-duality,gmpr3} it is also the M-theory origin of type IIA gauge supergravities. To obtain an M2-brane description associated to a massive supergravity in 10D, a decompactification limit is required to be taken. When performing it there are some properties that should be kept in the uplifted Hamiltonian in order to correspond to the massive supergravity sector and not to the massless one. For example, this limit should preserve finite couplings and in the case of Romans high energy description also the complex structure moduli. These facts cannot be obtained through ordinary decompactification limits as it happens when a 2-torus becomes a cylinder. Indeed, the formulation of the 11D supermembrane on a cylinder times 9D Minkowski corresponds to the 11D M2-brane on a trivial circle bundle whose low energy description corresponds to the maximal $N=2$ 10D type IIA supergravity. The massive supergravities (Romans and HLW) were expected to have a high energy description in terms of M-theory on nontrivial bundles \cite{hull-massive},\cite{lnr}. There are not many possibilities of implementing this idea with a single compact dimension. For the case of HLW gauged supergravity the explanation in terms of non trivial fiber bundle in the context of supergravity was done by \cite{llp1,llp2}.\newline
In this note we would like to discuss the existence of only two deformations of the 10D type II maximal supergravities, from the perspective of the M2-brane on nontrivial bundles with only one compact dimension.  We argue that only those M2-brane fiber bundles with parabolic monodromy in 9D are the ones who can be uplifted to M2-brane bundles whose low energy limit correspond to the 10D type II massive supergravities.
\section{Uplift of parabolic Monodromy M2-brane Torus bundles}
The M2-brane compactified on a torus in the Light Cone gauge compactified on $T^2\times M_9$ satisfies the following winding conditions
$$
\oint_{C_s}dX^1= 2\pi R_{11}(l_s+m_s Re(\tau)),\quad
\oint_{C_s}dX^2= 2\pi R_{11} m_s Im (\tau)
$$
where the torus is parametrized in terms of the  complex parameters $(R_{11}, \tau)$ and $(l_s,m_s)$ are the winding numbers which form a matrix over the torus target given by  $\mathbb{W}=\begin{pmatrix}l_1& l_2\\ m_1& m_2\end{pmatrix}$. Let us denote the complex one-closed one form $dX= dX^1+idX^2$, the supermembrane we are interested on, corresponds to the sector with nontrivial central charges i.e. those supermembranes satisfying the irreducible wrapping condition
\bea
\int_{\Sigma} dX\wedge \overline{dX}=n Area(\Sigma),\quad n\in\mathbb{Z}/ \{0\}
\eea
This condition is a quantization condition for a principal line bundle with first Chern class $c_1=n$.
 The Hamiltonian invariant section on a torus bundle wit parabolic monodromy (in the type IIB sector) and in complex coordinates corresponds to \cite{gmpr3},
\begin{equation}
\begin{aligned}
&H=\int_{\Sigma}d^2\sigma\sqrt{W(\sigma)}\left[\frac{1}{2}(\frac{P_{m}}{\sqrt{W}})^2+\frac{1}{2}(\frac{P\overline{P}}{W})+\frac{T^2}{4}\{X^ {m},X^{n}\}^2\right]\\&
+\left[\frac{T^2}{2}\mathcal{D}X^m\overline{\mathcal{D}}X^m+\frac{T^2}{8}\mathcal{F}\overline{\mathcal{F}}\right]\\
&-\int_{\Sigma} T^{2/3}\sqrt{W(\sigma)}\left[\overline{\Psi}\Gamma_{-}\Gamma_{m}\{X^{m},\Psi\}
+\frac{1}{2}\overline{\Psi}\Gamma_{-}\overline{\Gamma}\{X,\Psi\}\right]\\&
+\int_{\Sigma} T^{2/3}\sqrt{W(\sigma)}\left[\frac{1}{2}\overline{\Psi}\Gamma_{-}\Gamma\{\overline{X},\Psi\}\right]\\
& +\int_{\Sigma} \sqrt{W} \Lambda\left[ \frac{1}{2}\overline{\mathcal{D}}(\frac{P}{\sqrt{W}})+
\frac{1}{2}\mathcal{D}(\frac{\overline{P}}{\sqrt{W}})+\{X^m,\frac{P_m}{\sqrt{W}}\}-\{\Psi\Gamma_{-},\Psi\}\right].
\end{aligned}
\end{equation}
The constrain has been incorporated in the Hamiltonian through a lagrange multiplier $\Lambda$.
The symplectic covariant derivative is defined as,
\bea
\mathcal{D} \bullet=D\bullet+\{A,\bullet\dot\},\quad \mathcal{F}=D\overline{A}-\overline{D}A+\{A, \overline{A}\}
\eea
with $D=D_1+iD_2$ and $D_r \quad(r=1,2) $  as,
\bea
D_r\bullet=\frac{\epsilon^{ab}}{\sqrt{W}}2\pi R (l_r+m_r\tau)\theta_r^s\partial_a\widehat{X}^s \partial_b\bullet,
\eea
where $\theta$ is the set of residual discrete symmetries defined on the worldvolume. As it is explained in \cite{gmpr3} because of the construction, it depends on the parabolic monodromy $M_p$ as
$
\theta=(V^{-1}(M_p^*)^{-1}V)^{T},
$
where $(M_p^*)^{-1}=\begin{pmatrix} 1& -b_1\\ 0& 1\end{pmatrix}$
and $V\in SL(2,\mathbb{Z})=\begin{pmatrix} a & b\\ c& d\end{pmatrix}$ for $b_1=n b$ , being
$n= det \mathbb{W}$. Then, $\theta$ is also a parabolic matrix. The hamiltonian is constrained subject to first class constraints associated to the area preserving superdiffeomorphisms. The hamiltonian is T-dual invariant (locally and globally) \cite{joselen}, and it possesses a residual global symmetry associated to the parabolic subgroup of $SL(2,Z)$. The spectrum of the theory is purely discrete.

In \cite{khan} it was stated that for any fibre bundle $E$ with fiber $F$ base manifold $B$ and structure group $G$, the action of $G$ on $F$ produces a $\pi_0(G)$-action on the homology and cohomology of $F$. For $F$ being the 2-torus with structure group $G=Symp(T^2)$ it is shown that $\pi_0(G)\approx SL(2,\mathbb{Z})$. The $\pi_0(G)$-action has a natural action on $H_1(T^2)$ of $SL(2,\mathbb{Z})$ on a pair of integers $(a,b)\in \mathbb{Z}^2$ characterizing the $H_1(T^2)$ charges interpreted as the quantized Kaluza Klein momenta of the compactified supermembrane \cite{gmpr3}. It exists then a representation $M:\pi_1(B)\to\pi_0(G)$, and a bijective correspondence between the equivalence classes of symplectic torus bundles over $B$ inducing the module structure $\mathbb{Z}_{M}$ on $H_1(T^2)$ and the elements of the second cohomology group of $B$, $H^2(B,\mathbb{Z}_{M})$. The local coefficients $\mathbb{Z}_{M}$ run only over the integers allowed by the representation $M(n,m)$, see \cite{khan} for more details. We call this representation  $M$, \textit{Monodromy}.
%
We find that the natural monodromy that allows for decompactification such that the nontrivial behaviour will be codified in a single dimension -as happens in the case of the massive deformation uplift from nine to ten noncompact dimensions-, is the parabolic monodromy.
Given   $M_p$  a parabolic representation, then the coinvariant ,classes which classify the symplectic torus bundles with monodromy are in one to one correspondence with one integer in $\mathbb{Z}$. On the other side $\mathbb{Z}$ classify the principle line bundles over the base torus $B$. When decompactifying one dimension, in order to keep a massive deformation, it is natural to expect that some topological information must be preserved in the lifting. Since there is only one compactified dimension in 11D dimensions, it is natural to think that the topological classes will be related to the integers $\mathbb{Z}$. Additionally if we look at their coinvariant classes they are of the form.
\begin{equation}
C_{I}=\{Q+(g-1)\hat{Q}\}=\left( \begin{array}{c} \widehat{k}\\
q_0 \end{array} \right)
\end{equation}
where  $g\in M_p$, $\widehat{k}$ represents any arbitrary integer, and $q_0$ a particular one. For each $q_0\in \mathbb{Z}$ one has a coinvariant class,i.e, only one element of the homology basis  determines the coinvariant class. One of the circles of the target becomes then deedless in the process of decompactification. We then uplift the M2-brane torus bundle into a non trivial line bundle while the  members of the topological classes are preserved.
\section{Romans and the geometry of the twice-punctured 2-torus bundle}
The 10D Romans massive supergravity from the supermembrane bundle formulation perspective has to correspond to a non-trivial bundle.
Our proposal is to consider a M2-brane on a bundle with a twice punctured torus $\Sigma_{1,2}$ as the fiber (times $M_9$)\cite{gmr-romans}. The compactified manifold at the target is then a punctured torus. In this case in order to characterize the model we can use the Mandelstam map, \cite{mandelstam}, see also \cite{giddings}, to connect the punctured surface of the 2-torus with the Light Cone diagrams in order to characterize the moduli. The moduli of the twice-punctured torus is specified by four parameters $(R,\tau,Z_1,Z_2)$. Considering the conformal structures (that is neglecting just for a moment the scaling parameter $R$) the torus can be conformally mapped to the Light Cone diagrams as originally shown by \cite{mandelstam} and laterly developed by \cite{giddings} in the aim of showing the equivalence between Polyakov and Light Cone bosonic string formulation. The moduli of the Light cone diagrams corresponds to those of an infinite long cylinder with one loop inside. The borders the cylinder corresponds to the singularities associated to the punctures at infinity with residues $\pm 1$. As explained by \cite{giddings} its moduli is specified by four parameters: $(t,\alpha, \theta_1,\theta_2)$, where $0<t<\infty$  represents the interaction time between the different closed strings, $0<\theta<2\pi$ are the twist angles on the internal cylinders, $\alpha$ represents the internal momentum fraction, i.e. the sum of the positive residues $\sum_i \alpha^+_i=n$\cite{mr}. These parameters can be mapped in the ones of the torus $(\tau,Z_1,Z_2)$ being $\tau$ the complex modular parameter of the torus, and $(Z_1,Z_2)$ the positions of the punctures on the Riemann surface $\Sigma_{1,2}$.
by mean of the following identifications:
\bea
2\pi i (Z_1-Z_2)= (\theta_1+\theta_2)\alpha_1-\theta_2
-2\pi i \tau\eea
However to characterize individually those moduli is necessary to know the abelian integral defined as follows \cite{mandelstam} $w=\int^{z}\omega$, with $w=x+iy$ identified with the parameters of the string $w=t+i\sigma$. Locally $\omega=dw$
\begin{equation}
\begin{aligned}
w=&ln\theta[\begin{pmatrix} 1\\1\end{pmatrix}](z-\xi_1,\tau)-ln\theta[\begin{pmatrix} 1\\1\end{pmatrix}](z-\xi_2,\tau)\\ &+2\pi i z\frac{Re(2\pi i (\xi_2-\xi_1)}{Re(2\pi i \tau)}
\end{aligned}
\end{equation}
where $\xi_1,\xi_2$ indicate the position of the simple zeroes of $\frac{dw}{dz}.$
This geometrical structure when applied to the M2-brane \cite{gmr-romans}, fits very well in all the prescribed requirements: it corresponds the decompactification limit of a torus bundle times 9D Minkowsky target space into a twice-punctured torus bundle times 9D Minkowski space time which effectively is a non trivial compactification associated to a $(1,1)$-Knot space in 10 noncompact dimensions. Indeed,  it corresponds to a massive supermembrane \cite{gmr-romans}. We believe this topology could be related to the massive branes analysis \cite{blo}. In that study the 11D supergravity is recovered from Romans one by means of a mass term which depends on a killing vector present in the 11D space formulation. The analysis is non covariant, and we find it natural from the point of view of the type of the decompactification manifold we are considering.
\section{Global description of the M2-brane on the twice punctured torus bundle.}
Globally the M2-brane nontrivial bundle corresponds to a fiber bundle, whose base corresponds to the Riemann surface of the M2-brane worldvolume foliation -for simplicity we choose to be also a 2-torus- and an associated fiber corresponding to the target space $M_9\times \Sigma_{1,2}$ with 10 non-compact dimensions times a compact one. This fiber bundle should be understood as the decompactification of the M2-brane on the symplectic torus bundle with parabolic monodromy. The mapping class group of the twice punctured torus bundle correpsonds to a $(1,1)$ knot \cite{cattabriga}. There is a residual monodromy associated to the twice punctured torus bundle which is inherited from the parabolic monodromy, indeed the two punctures that lead to a noncompact distinguished 10-th dimension, are kept invariant under the parabolic monodromy. More formally and following \cite{cattabriga} let be $H(F_g,P)$ the homeomorphisms group preserving the orientation, $F_g$ a fiber of genus $g$ and $P$ the punctures, such that $h(P)=P$. The punctures are mapped into punctures though not necessarily become fixed. Let define the isotopy group of $H$, it corresponds to the mapping class group of the twice-punctured two torus $\Sigma_{1,2}$,
$
MCG(\Sigma_{1,2})=\Pi_0(H(\Sigma_{g},P))
$
then $MCG(\Sigma_{1,2})\equiv PMCG(\Sigma_{1,2})\oplus Z_2$
with $PMCG(\Sigma_{1,2})$ being the subgroup generated by the three Dehn twists $(t_{\alpha},t_{\beta},t_{\gamma})$ associated to the homology basis $\alpha, \beta,\gamma$. In the case of the compact two-torus  $T^2$ the two Dhen twwist generators associated to the one-cycles between the two punctures  are identified $t_{\beta}=t_{\gamma}$. Indeed there exists an epimorphisms $\Omega$ between the mapping class group of the compact 2-torus $\Sigma_1$  $MCG(T^2)$ and the Pure Mapping Class group (preserving punctures),
$$
\Omega: PMCG(T^2)\to MCG(T^2)= SL(2,Z)
$$
\bea
\Omega(t_{\alpha})=\begin{pmatrix} 1 &0\\ 1& 1\end{pmatrix},\quad \Omega(t_{\beta})= \Omega(t_{\gamma})=\begin{pmatrix} 1 &-1\\ 0& 1\end{pmatrix}
\eea
These two parabolic matrices through multiplication generate the full $SL(2,Z)$. The monodromy of the punctured torus bundle over a 2-torus is defined in terms of the symplectic parabolic torus bundle over a torus as follows:
let consider $a,b$ two integers classifying the cohomology basis of the base torus $\Sigma$ the monodromy is defined as
$
M_{10D}: \Pi_1(\Sigma_1)\to \Pi_0(\Sigma(1,2))
$
then the 9D parabolic monodromy is is uplifted through decompactification to the 10D monodromy $M_{10D}$,
\bea M_{9D}=\begin{pmatrix} 1 &1\\ 0& 1\end{pmatrix}^{a+b}\to\quad M_{10D}=t_{\beta}^{a+b}.\eea
\section{Conclusions}
M2-branes formulated on symplectic torus bundles with monodromy in $SL(2,Z)$ are associated at low energies with the 9D type II gauged supergravities. Only those bundles which have parabolic monodromy can be uplifted to ten dimensions keeping a residual topological invariant associated to the massive deformation of supergravity theories. There are two possible M2-brane bundles with parabolic monodromies of the type IIA sector, linearly and non-linearly realized, associated at low energies to Romans and HLW supergravities in 10D. We consider that Romans supergravity is the low energy description of the 11D  M2-brane on a twice punctured 2-torus bundle. The 10-th dimension is noncompact but distinguished, there is only one compact dimension, consequently it does not admit a $M_{10}$ target-space description. We conjecture that the non-trivial topology origin in 11D could be related to the 11D non-covariant cosmological term of 11D Supergravity found in \cite{blo}. A deeper study is required  to give a proper answer to this question.
\section{Acknowledgements} The authors thank A.Tomassiello, Y. Lozano, P. Meessen, T. Ortin and Diego Regalado for interesting comments and discussions at different stages of this research. MPGM is supported by Mecesup ANT1398, Universidad de Antofagasta, (Chile). A.R. is partially supported by Projects Fondecyt 1121103 (Chile). MPGM and AR participate in the EU-COST Action MP1210 `The String Theory Universe'.

\end{document}